\documentclass[preprint]{aastex}
\usepackage{graphicx}

\begin{document}
\slugcomment{To appear in {\it Astrophysics of Dust} (2004)}
\title{Interstellar Extinction in the Milky Way Galaxy}
\author{Edward L. Fitzpatrick}
\affil{Villanova University, 800 Lancaster Avenue, Villanova, PA 19085}

\begin{abstract}
I review the basic properties of interstellar extinction in the Milky
Way galaxy, focusing primarily on the wavelength dependence within the
IR through UV spectral region.  My primary goal is to review the
evidence supporting the idea that Galactic extinction curves can be
considered a 1-parameter family characterized by their value of $R_V
\equiv A_V/E(B-V)$.  Based on analysis of new (i.e., {\em 2MASS}) and
old (i.e., {\em IUE}) data for $\sim$100 sightlines, I show that the
UV, optical, and IR wavelength regimes do display coherent variations,
but with too much intrinsic scatter to be considered truly correlated.
A 1-parameter family can be constructed which illustrates these broad
trends, but very few individual sightlines are actually well-reproduced
by such a family and disagreement with the mean trends is not a
sufficient condition for ``peculiarity.'' Only a very small number of
extinction sightlines stand out as truly peculiar.  It is likely that
simple variations in the mean grain size from sightline to sightline
are responsible for much of the coherent variability seen in Galactic
extinction, and might also explain the ``peculiar'' extinction
long-noted in the Magellanic Clouds.
\end{abstract}

\section{Introduction}

This conference is a testament to the wide ranging impact of dust
grains on the properties of the interstellar medium and to the rich
variety of research opportunities in the interstellar dust field.
Among these studies, interstellar extinction --- the absorption and
scattering of light by grains --- often plays a central role. The
ability of grains to transmit, redirect, and transmute electromagnetic
energy is enormously important to the physics of interstellar space.
The wavelength dependence of extinction provides important diagnostic
information about the physical properties of grain populations, and
often serves as a first test for possible grain models.  And, perhaps
most far-reaching, interstellar extinction profoundly limits our
ability to study the universe.  It is certainly fair to say that many
more astronomers care about extinction, than care about the dust grains
that produce it!

In recent history there have been a number of excellent review articles
about, or including, the basic properties of interstellar extinction in
the Milky Way galaxy (e.g., Savage \& Mathis 1979; Massa \& Savage
1989; Draine 1995; and Draine 2003).  In addition, the book by Whittet
(2003) provides a valuable general reference.  Since the pace of
research in the extinction field is such that the astronomical
community is not really demanding a new review article every year, I
have taken a liberal approach to the notion of a ``review.''  In this
paper, my aim is principally to review an idea, rather than a field.
The idea is that --- despite the wide range of spatial variations
present --- the $\lambda$-dependence of Galactic extinction can be
viewed as a smoothly varying 1-parameter family and that the normalcy
of an extinction sightline can be judged by how well it fits into this
family.  In a sense, this article is a natural follow-up to that by
Massa \& Savage (1989) presented at the Santa Clara dust meeting.
Massa \& Savage gave a comprehensive review of the detailed
$\lambda$-dependence of extinction and included a discussion of the
then-recent results of Cardelli, Clayton, \& Mathis (1988, 1989) who
first showed a relationship between extinction properties in the
IR/optical and in the UV.  It is these results which have given rise to
the 1-parameter family scenario.

I begin this article with a brief description of the basic properties
and terminology of Galactic extinction (\S 2) and then give a short
history of the key developments which provide the primary background
for this paper (\S 3).  In \S 4, I will examine the evidence for a
multi-wavelength connection between IR, optical, and UV extinction,
i.e., the issue of a 1-parameter family, based on a new analysis of UV,
optical, and IR data. The analysis incorporates a threefold increase
over past studies in the number of sightlines for which IR and UV data
are combined, and utilizes a new technique for deriving extinction
curves which promises to allow precise extinction studies to be
extended to more lightly-reddened sightlines than previously possible.
Finally, in \S 5 I will revert to classical``review-mode'' and note a
number of recent programs and results which are helping to clarify our
view of the $\lambda$-dependence of Galactic extinction and the
relationship between extinction properties and interstellar
environment.

\section{The Extinction Basics}

While the ultimate interpretation of results from extinction studies
may be very complex, the measurements themselves for Galactic (and many
extra-Galactic) sightlines are, in principle, straightforward.  The
most commonly used technique for deriving the $\lambda$-dependence of
extinction, the ``pair method,'' is illustrated in idealized form in
Figure 1.  Measuring the extinction produced by an interstellar cloud
requires photometric or spectrophotometric observations of the spectral
energy distribution (SED) of some astronomical object --- usually a
star --- located behind the cloud and identical measurements of an
identical object located at an identical distance, but unaffected by
interstellar dust.  The total extinction in magnitudes at each
wavelength $A_{\lambda}$ (which is essentially a measure of the optical
depth $\tau_{\lambda}$) is then simply computed from the ratios of the
SEDs as shown in Figure 1.  Note that the assumption implicit in this
discussion is that the star is located far from the interstellar cloud
(as indicated in Figure 1) and that no photons are scattered {\em into}
the line of sight.  Such an assumption is not necessarily valid in the
case of extinction produced by circumstellar material.
\begin{figure}
\epsscale{1.00}
\plotone{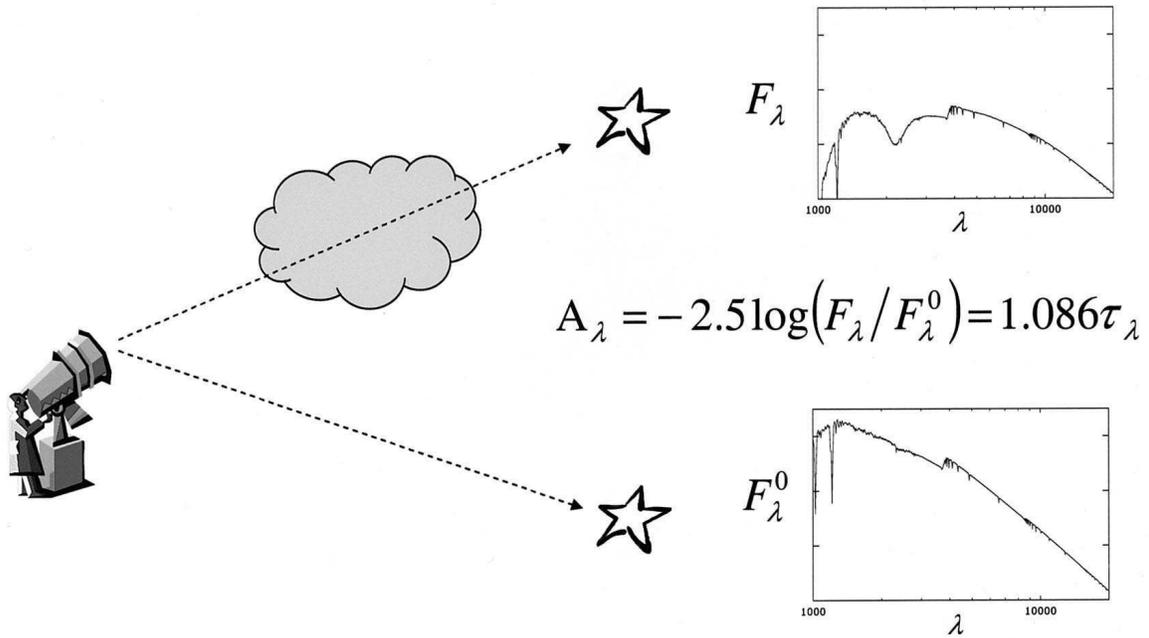}
\caption{A schematic illustration of the ``Pair Method,'' the principal technique used to study Milky Way extinction.}
\end{figure}

Because reddened/unreddened star pairs are nearly never at the same
distance and, in fact, the stellar distances are usually poorly
determined, the total extinction $A_{\lambda}$ is rarely computed
directly.  Instead, the stellar SEDs are usually normalized by the
flux in a common wavelength region before computing the extinction.
The usual choice for this is the optical $V$ band.  This normalized
extinction $A_{\lambda-V}$ (computed by substituting $F_\lambda/F_V$
and $F_\lambda^0/F_V^0$ into the equation in Figure 1) is also known
as the color excess $E(\lambda-V)$ and is related to the total
extinction by $E(\lambda-V)$ = $A_{\lambda} - A_V$.

To effectively compare the $\lambda$-dependence of extinction among
sightlines with very different quantities of interstellar dust, the
color excess $E(\lambda-V)$ itself must be normalized by some
factor related to the amount of dust sampled. The optical color excess
$E(B-V)$ is usually employed for this purpose.  This second
normalization yields the most commonly found  form for observed
``extinction curves'', namely $E(\lambda-V)/E(B-V)$.  This normalized
extinction is related to the total extinction $A_{\lambda}$ through the
relation $E(\lambda-V)/E(B-V) = A_{\lambda}/E(B-V) - A_V/E(B-V)$.  The
quantity $A_V/E(B-V)$, i.e., the ratio of total extinction to color
excess in the optical region, is usually denoted $R_V$.  If its value
can be determined for a line of sight, then the easily-measured
normalized extinction can be converted into total extinction.

It has been noted often that $E(B-V)$ is a less-than-ideal
normalization factor.  Certainly a physically unambiguous quantity,
such as the dust mass column density, would be preferred, or even a
measure of the total extinction at some particular wavelength, such as
$A_V$.  However, the issue is simply measurability.  We have no
model-independent ways to assess dust mass and total extinction
requires either that we have precise stellar distances or can measure
the stellar SEDs in the far-IR where extinction is negligible.  While IR
photometry is now available for many stars through the
{\em 2MASS} survey, the determination of total extinction from these data
still requires assumptions about the $\lambda$-dependence of extinction
longward of ~2$\mu$m and can be compromised by emission or scattering
by dust grains near the stars.  In this paper, all the observed
extinction curves will be presented in the standard form of
$E(\lambda-V)/E(B-V)$.  Only in the case of model curves, and for
illustrative purposes, will alternate normalizations be employed.

The quality of a pair method extinction curve clearly depends on how
well a reddened object can be matched with an unreddened comparison
object.  Massa et al. (1983) discuss the effects and magnitude of
``mismatch errors'' on extinction curves.  In general, mismatch effects
become less important as the amount of extinction increases, and
strongly limit our ability to probe the $\lambda$-dependence of
extinction in lightly reddened sightlines, such as through the nearby
interstellar medium and the galactic halo.  In \S 4, I will present
results based on the substitution of stellar atmosphere models for the
unreddened member of a pair method star team. Along with a number of
other advantages, this technique promises to greatly improve the
accuracy of extinction curves derived for low-reddening sightlines.

Several decades of work utilizing pair method extinction curves
spanning the near-IR through UV spectral domain have provided a good
estimate of the ``average'' $\lambda$-dependence of extinction in the
Milky Way galaxy.  (I will defer describing what I mean by ``average''
until \S 3 below).  Figure 2 shows the familiar shape of this ``Average
Galactic Extinction Curve'', plotted
as $E(\lambda-V)/E(B-V)$ vs. inverse wavelength.  This type of
presentation was clearly invented by UV astronomers to emphasize that
spectral region, but also has the advantage of being essentially a plot
of normalized extinction cross section vs.  photon energy.  The
Average Curve itself is shown by the thick solid and dashed curve, with
a number of specific features labeled.  I use the solid curve to denote
the spectral regions where I think the form of the extinction is
particularly well-determined.  These are the IR 1.25--2.2~$\mu$m region
(based on {\em 2MASS JHK} photometry), the optical 5500-3600~\AA\/
region (based on {\em UBV} and {\em uvby} photometry) and the UV
2700--1150~\AA\/ region (based on {\em IUE} satellite
spectrophotometry).
\begin{figure}
\plotone{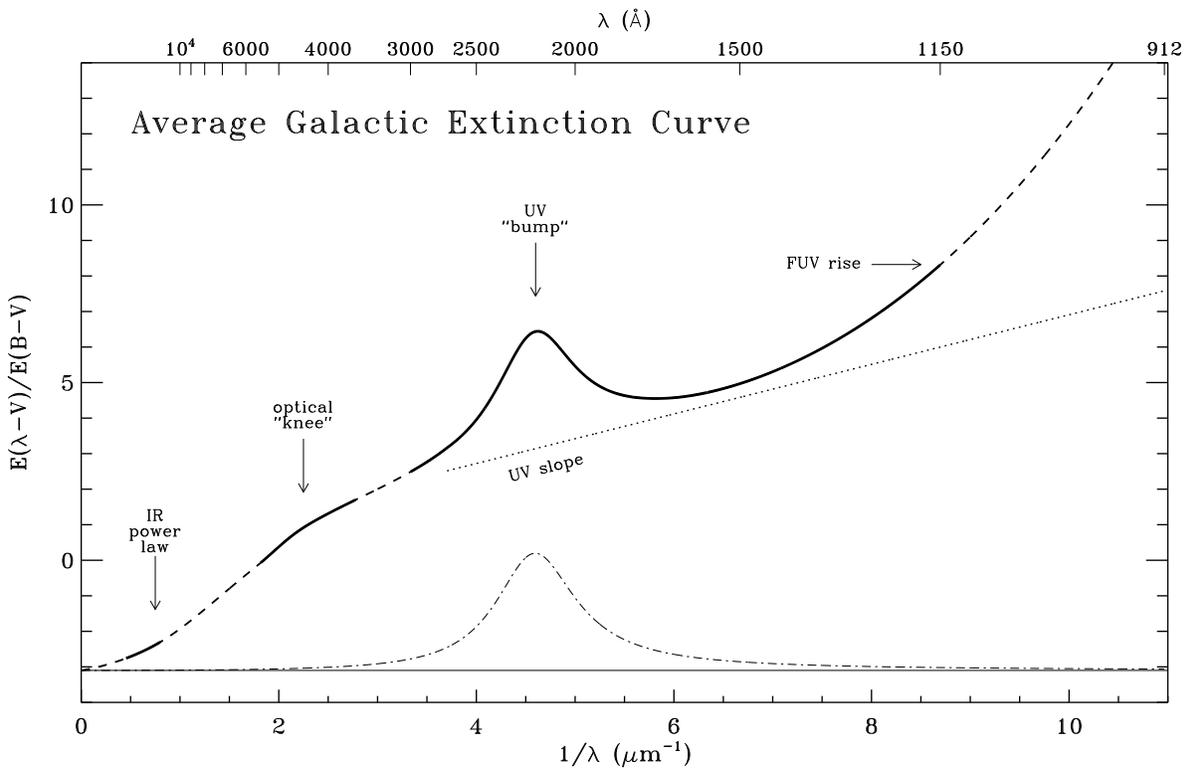}
\caption{The average Milky Way extinction curve, corresponding to the
case $R_V = 3.1$, as computed by Fitzpatrick 1999.  See the discussion
in \S 2.}
\end{figure}

The most prominent feature in the curve is the ``bump'' at ~2175 \AA.
This has been shown to be a pure absorption feature with a
Lorentzian-like cross section, and likely arises from a specific
physical process occurring on a specific --- although currently
unidentified --- type of dust grain (see \S 3).  The mean profile of
the bump is shown by the dash-dot curve at the bottom of the figure.
If the profile of the bump is removed from the extinction curve, the
underlying extinction in its vicinity is seen to be linear and its
slope can be used to characterize the steepness of UV extinction (``UV
slope'' in Figure 2).  At wavelengths shortward of $\sim$1500 \AA\/
the extinction departs from the linear extrapolation, rising more
rapidly and generally with pronounced curvature (``FUV rise'' in Figure
2).

Probably the most outstanding observational characteristic of
interstellar extinction is its spatial variability.  I show this
graphically in Figure 3 where analytic fits to 96 Galactic extinction
curves are overplotted, in the form $E(\lambda-V)/E(B-V)$. Only the
highly constrained IR, optical, and UV spectral regions are shown.
These curves were produced as part of the analysis discussed in \S 4
below, and their details will be discussed there.  Here they simply
serve to demonstrate the degree of variability found along Milky Way
sightlines.  This variability is a two-edged sword.  It can provide
endless misery to those who seek to correct astronomical observations
for the effects of extinction, but also a wealth of data for those who
model the interstellar dust grain populations.  The extinction
variations presumably reflect general differences in the grain
populations from sightline to sightline.  Understanding how the various
spectral regions of the curves relate to each other and how they
respond to changes in the interstellar environments can provide
information critical for characterizing interstellar grains.
\begin{figure}
\plotone{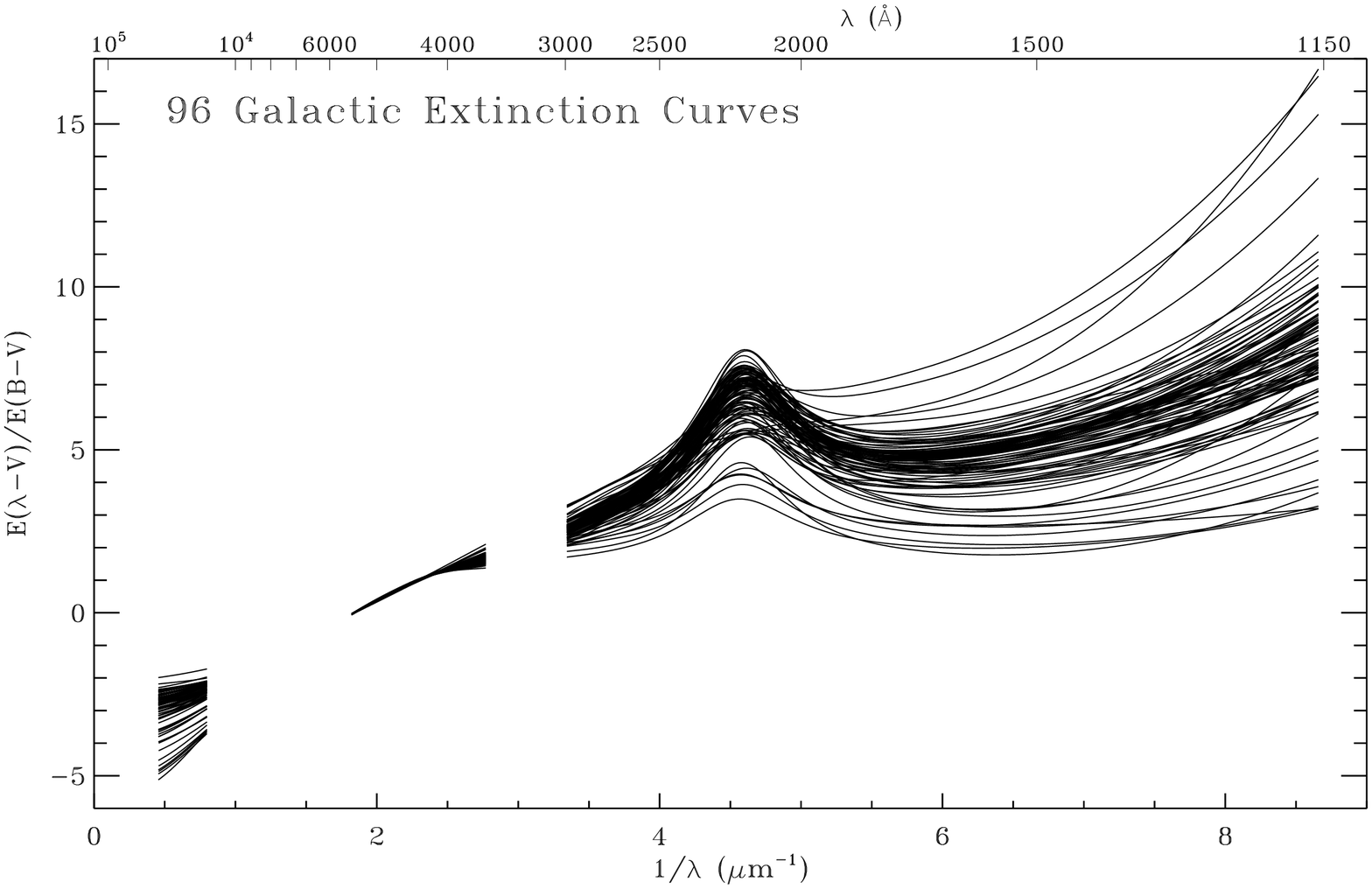}
\caption{Analytical representations of the 96 IR-through-UV extinction
curves to be discussed in \S 4.  These illustrate the wide range of
extinction properties observed in the Milky Way.}
\end{figure}

\section{A Brief History}

A detailed history of the development of our understanding of
interstellar extinction is beyond the limited scope I have set for this
article.  In this section I will outline only a few key items which
play a role in the re-analysis of Galactic extinction data to be
presented in \S 4.

The $\lambda$-dependence of IR and optical extinction has been studied
extensively using ground-based photometry and sightline-to-sightline
variations have long been recognized (e.g., Johnson 1965).  These
variations can be characterized by their $R_V$ values, which range from
$\sim$2 to $\sim$5.5 for sightlines through the diffuse ISM, with a
mean value of $\sim$3.1. The shape of the extinction curve at
wavelengths longward of $\sim$1.2 $\mu$m (i.e., the $J$ band) has been
suggested as invariant and universal, following a power law dependence
proportional to $\lambda^{-1.84}$ (see the discussion in Whittet
1992).

UV extinction measurements require space-based observations and date
back to the late 1960's. The 2175 \AA\/ bump was discovered by Stecher
(1969) and early observations clearly revealed the presence of
sightline-to-sightline variations in the $\lambda$-dependence of UV
extinction.  Soon after, measurements of the UV component of the
diffuse galactic light revealed a broad minimum in the scattering albedo
centered near 2200 \AA, indicating the bump to be an absorption feature
(Witt \& Lillie 1973). The large spectrophotometric database
accumulated by the {\em IUE} satellite over its long operational
lifetime (1978-1996) has provided the best resource for studying UV
extinction (see references in Fitzpatrick 1999) and these data are the
source of the UV curves shown in Figure 3.

Despite the daunting degree of variation present in Figure 3, the
extinction curves do follow some rules and do not assume arbitrary
appearance.  Savage (1975) first demonstrated that the shape of the 2175
\AA\/ bump is well represented by a Lorentzian function.  Fitzpatrick
and Massa (1986) later showed that, although the strength and width of
the bump vary markedly in the ISM, it always retains its
Lorentzian-like form and has a nearly invariant central peak position.
Fitzpatrick and Massa (1986, 1988, and 1990) also showed that the whole
range of known UV extinction curves could be reproduced extremely well
by a single analytical expression with a small number of free
parameters.  This expression consists of (1) a Lorentzian-like bump
term (requiring three parameters, corresponding to bump FWHM $\gamma$,
position $x_0$, and strength $c_3$), (2) a linear term underlying the
bump and extrapolated into the far-UV (two parameters, the intercept
$c_1$ and slope $c_2$), and (3) a far-UV curvature term (one parameter,
a scale factor $c_4$).  As far as I know, it is still true that {\em
all} UV extinction curves measured so far can be reproduced with this
parameterization scheme to a level consistent with observational error.

Cardelli, Clayton, \& Mathis (1988, 1989; and see also Mathis \&
Cardelli 1992) were the first to demonstrate a link between UV
extinction and that in the optical/IR region by showing that $R_V$
correlates with the level of UV extinction.  Essentially, sightlines
with large $R_V$ values tend to have low UV extinction, and vice
versa.  These results were important in demonstrating a degree of
coherent behavior over the full wavelength range observed for
interstellar extinction and for providing a simple recipe for
determining a meaningful average extinction curve.  I.e., since it is
well-established that the mean value of $R_V$ in the diffuse ISM is
$\sim$3.1, it is reasonable to define the Average Galactic Extinction
Curve as that which corresponds to $R = 3.1$.  This is how the curve
in Figure 2 was defined.  

A comparison of the Cardelli et al. $R_V$ values with the UV extinction
curve parameters of Fitzpatrick \& Massa suggests that --- while other
effects are present --- the primary basis of the correlation discovered
by Cardelli et al. is a relationship between 1/$R_V$ and the UV slope
component (see also Jenniskens \& Greenberg 1993).  Thus the general
$R_V$-dependence of extinction can be distilled to the simple statement
that when extinction curves are intrinsically steep in the optical
(i.e., large 1/$R_V$) they remain steep in the UV (large UV slope), and
vice versa.  Based on this, Fitzpatrick (1999) produced a much
simplified formulation of the $R_V$-dependence which reproduced the
main effects seen by Cardelli et al.

\section{A New Look At Old Data or ``Is Galactic Extinction Really a 1-Parameter Family?''}

Since the early 1990s, there has been very little new data to advance
our basic understanding of the multi-wavelength behavior of Galactic
interstellar extinction. The {\em IUE} database was
complete by then, optical photometry had already been available for all stars
with UV extinction curves, and Cardelli et al. had already exploited
virtually all the IR data available for stars with UV extinction
curves.  This latter point is significant since the Cardelli et al.
results are based on data for only 29 stars, out of the several hundred
for which UV extinction curves might be derived.  What {\em has}
changed over the last 10 years, in my view, is the {\em perception} of
the $R_V$-dependence discovered for Galactic extinction curves.
Increasingly, Galactic extinction curves are referred to as a
``1-parameter family'' (with $R_V$ as the parameter) and the degree to
which a given curve agrees with Cardelli et al.'s (or Fitzpatrick's
1999) enunciation of the $R_V$-dependence used as a test of its
normalcy.

The time is now right to reexamine the whole issue of the connection
between IR, optical and UV extinction.  The primary driver for this is
the recent All-Sky Data Release from the {\em 2MASS JHK} photometric
survey, which allows the determination of $R_V$ for nearly all the
stars in the UV extinction database.  A secondary motivation for me is
the recent development by Derck Massa and myself of a technique for
utilizing stellar atmosphere models to derive extinction curves for
reddened stars, rather than relying on unreddened standard stars.  The
technique is described by Fitzpatrick \& Massa (1999) and had been
utilized in a number of studies of eclipsing binary stars in the Large
Magellanic Cloud (see Fitzpatrick et al. 2003 and references within).
The benefits to this technique are numerous, including: 1) the
elimination of the subjective process of determining the pair method
standard; 2) a large reduction in mismatch error in the resultant
extinction curves; 3) the ability to determine reliable extinction
curves for lower E(B-V) sightlines than possible previously; and 4) the
ability to determine reliable UV extinction curves for later type stars
(i.e., as cool as early-A) than previously possible.  The principal
drawback is that the stars to which the technique can be applied must
be restricted to the spectral domain in which the models have been
shown to precisely and accurately reproduce the stellar SEDs.
Currently this forces us to consider only main sequence stars in the
late-O through early-A range.

\subsection{The Data}

Massa and I are currently producing near-IR through UV extinction
curves for the several hundred stars in the {\em IUE} database which
are amenable to our technique and a full description of the results is
being prepared.  In this article, I present the first results from a
core sample of 96 stars.  These consist of the stars in the atlas of
Fitzpatrick \& Massa (1990), plus the study by Clayton \& Fitzpatrick
(1987) of the Trumpler 37 region, plus a number of other sightlines not
part of either of those studies, including HD27778, HD29647, HD294264,
Brun 885, HD38023, HD38051, HD62542, HD210072, and HD210121.  Several O
stars from the Fitzpatrick \& Massa (1990) sample were eliminated from
consideration (HD48099, HD167771, HD199579, HD229296, Trumpler 14 \#20,
and CPD--59 2600) because their strongly developed C IV stellar wind
lines suggested that the hydrostatic stellar atmosphere models being
utilized were likely inappropriate.

The analysis technique is described in detail in the references listed
above.  In brief, it consists of modeling the observed stellar SEDs by
combining model atmosphere fluxes with a flexible description of the
shape of the IR-through-UV extinction curve.  The output of the fit
includes the stellar properties (potentially $T_{eff}$, $\log g$,
[m/H], and $v_{microturb}$) and the set of parameters which describe
the shape of the extinction curve.  In the IR, at wavelengths longward
of 1 $\mu$m, we assume the extinction to follow a power law form given
by $A_{\lambda}/E(B-V) = K \times \lambda^{-\alpha}$, where both the
scale factor K and the exponent $\alpha$ may be determined by the fit.
In the UV, at wavelengths shortward of 2700 \AA\/, we adopt the
6-parameter fitting function from Fitzpatrick \& Massa (1990) described
in \S 3 above. All 6 parameters are determined by the fit. The UV and
IR regions are linked by a cubic spline interpolation passing through
optical anchor points located near the $V$, $B$, and $U$ effective
wavelengths.  The values of these 3 spline points are strongly
constrained by the optical photometry, the value of $R_V$, and the
normalization imposed on the curves, and are all determined by the fit,
along with $R_V$ itself.  The 6-parameter UV fitting function is
restricted to $\lambda < 2700$ \AA\/ because the assumption of a linear
component underlying the bump begins to breakdown towards longer
wavelengths.  This effect can be seen (although it was not recognized
at the time) in Figure 2 of Fitzpatrick \& Massa (1988).
    
Our primary datasets are {\em IUE} spectrophotometry in the UV, {\em
2MASS} photometry in the IR, and {\em UBV} and $uvby\beta$ photometry
in the optical.  We use Kurucz's ATLAS9 atmosphere models to represent
the surface fluxes of the B stars and the Lanz \& Hubeny (2003) non-LTE
models for the O stars.

Figure 4 shows examples of best-fitting SED models for five of the
stars in the current sample.  The fitting procedure is clearly capable
of reproducing the observed SEDs at a level consistent with
observational error.  Figure 5 shows the extinction curves derived from
the SED fits for 11 of the stars, including those in Figure 4.  (The
parameterized versions of whole set of 96 extinction curves were
already shown in Figure 3.)  The full set of SED fitting results and
extinction curves will be published by Fitzpatrick \& Massa (in
preparation).
\begin{figure}
\epsscale{0.80}
\plotone{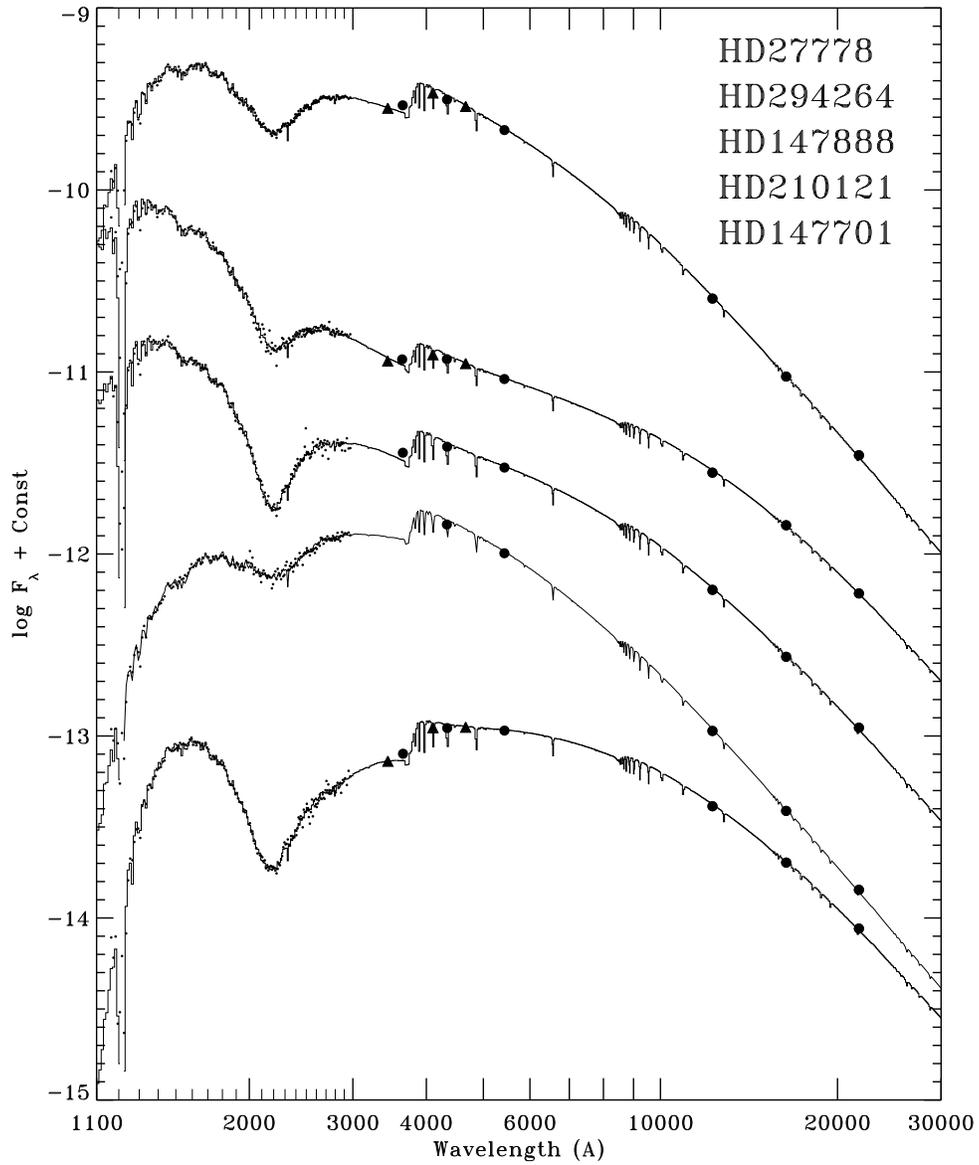}
\caption{Five examples of the fits to reddened star SEDs. IR data are
{\em 2MASS} photometry, optical data are {\em UBV} (circles) and {\em
uvb} (triangles) photometry, and UV data are {\em IUE}
spectrophotometry.  The solid curves are the best-fitting stellar
atmosphere models reddened by extinction curves whose form is
determined by the fitting procedure.  The SEDs have been shifted
vertically for clarity.}
\end{figure}
\begin{figure}
\epsscale{0.80}
\plotone{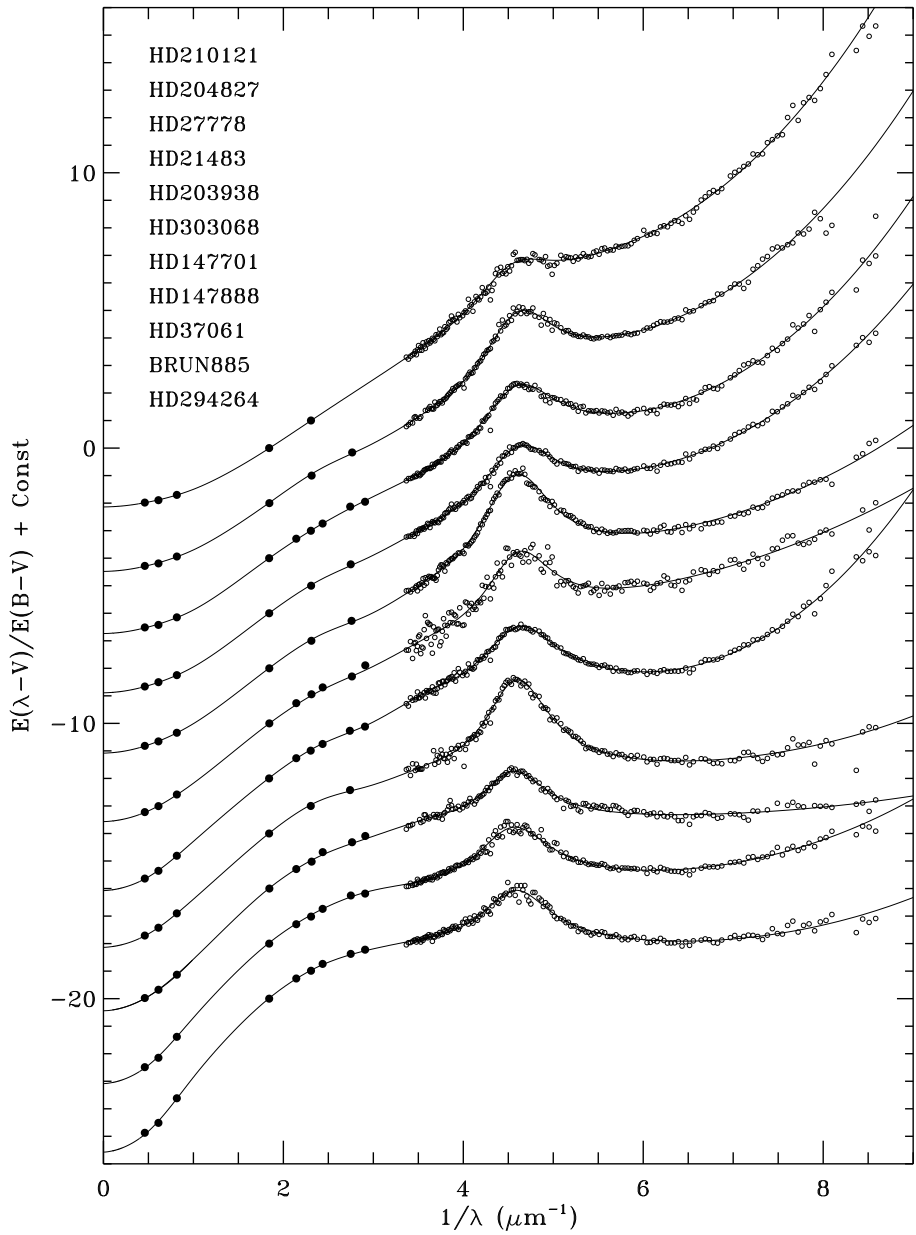}
\caption{Examples of extinction curves produced with the SED fitting
procedure. The smooth curves are the analytical representations of the
extinction, which are determined by the fits.  The data points show the
actual ratios between the reddened star SEDs and the model atmosphere
calculations. The curves have been successively offset downwards by 2
units for clarity.  The data point at $V$ ($1/\lambda = 1.83$
$\mu$m$^{-1}$) should be located at $E(\lambda-V)/E(B-V) = 0$ for each
curve.}
\end{figure}

\subsection{The Extinction Results}

The primary results for the 96-star core sample are illustrated in
Figures 6 through 11.  While a full analysis of these data is yet to be
performed, the major highlights can be summarized briefly:\\

$\bullet$ The IR extinction (i.e., at $\lambda$ $>$ $\sim$1 $\mu$m) is
well-represented by a power law of the form \[ A_{\lambda}/E(B-V) = k\times\lambda^{-\alpha} \]
with a universal exponent $\alpha = 1.84$ as
suggested by Whittet (1992). The IR scale factor $k$ varies smoothly
and linearly with $R_V$, which ranges from $\sim$2.1 to $\sim$5.6 in
the current sample. See Figure 6.
\begin{figure}
\plottwo{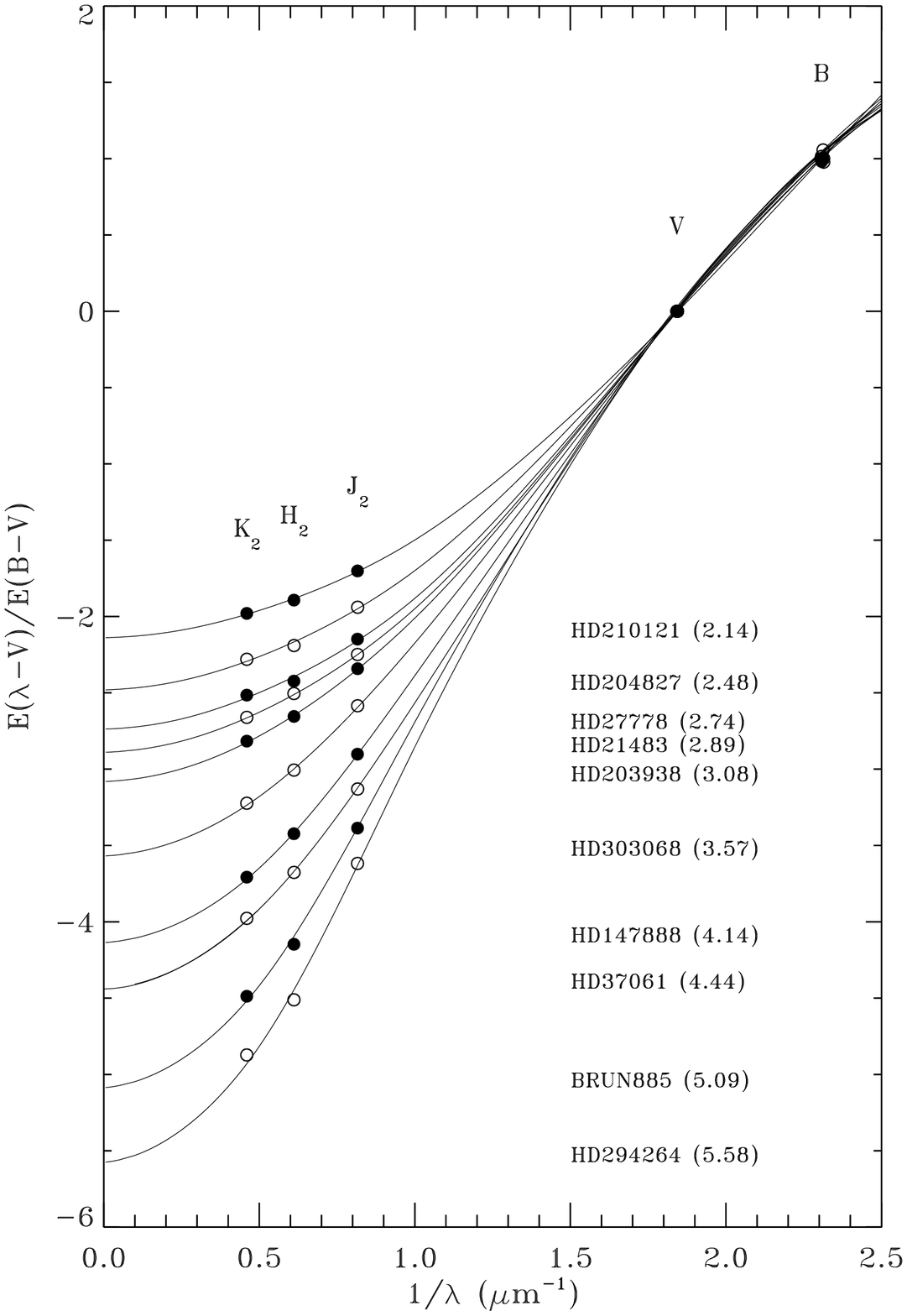}{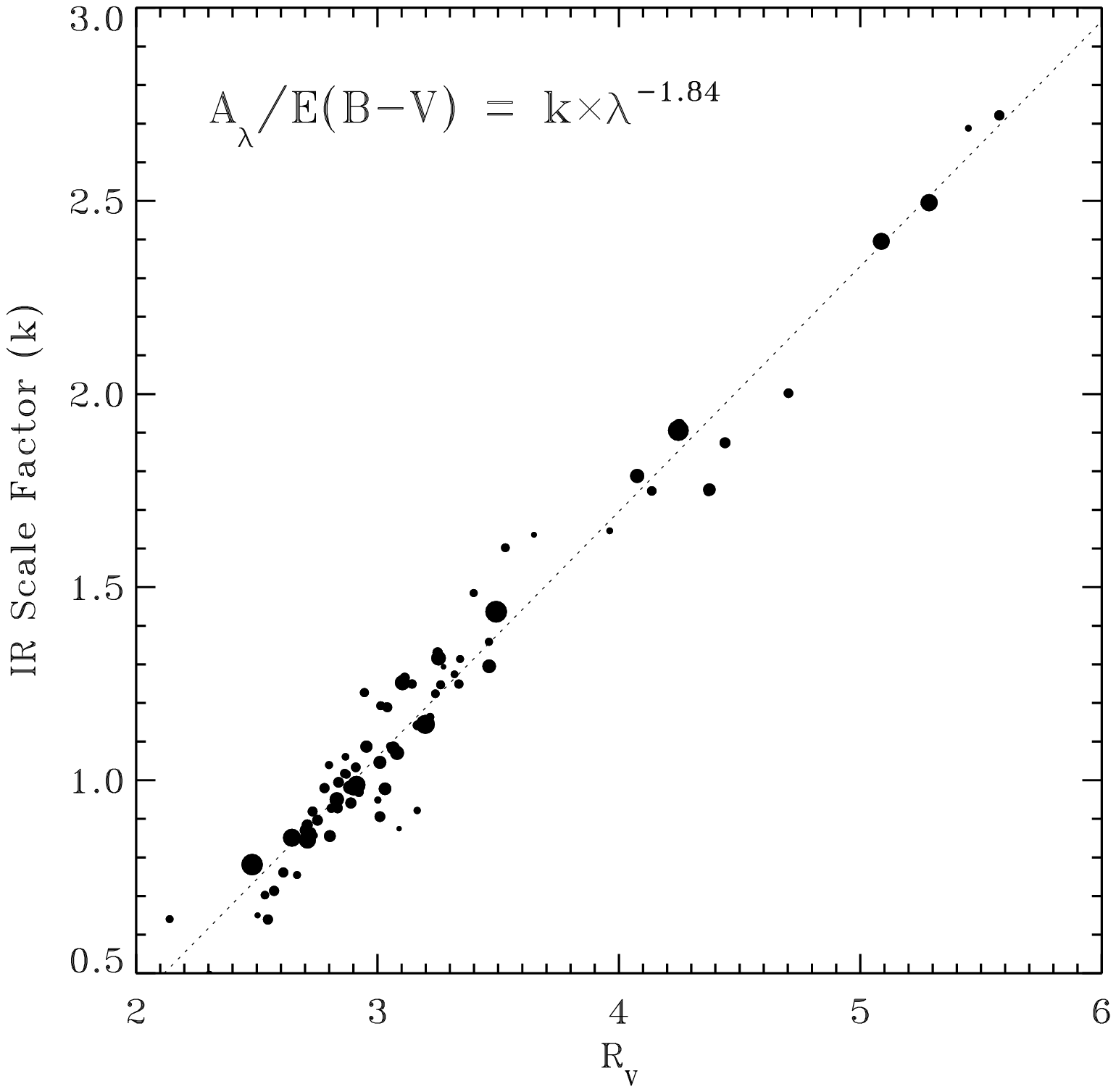} 
\caption{The lefthand panel shows the IR/optical extinction curves for
10 stars in the core sample.  The stars and their $R_V$ values are
identified in the figure.  The IR portion of the curves consist of
power laws of the form $A_{\lambda}/E(B-V) = k \times\lambda^{-1.84}$
smoothly joined to the optical region with a cubic spline
interpolation. The righthand panel shows that the IR scale factor $k$
is a simple and well-correlated function of $R_V$. The dotted line is a
linear least-squares fit corresponding to $k = 0.63 R_V - 0.84$. The
size of the symbols in the panel is proportional to the {\em E(B--V)}
of the sightline. Internal measurement errors are typically $\pm$0.03
for $k$ and less than $\pm$0.1 for $R_V$.}
\end{figure}

$\bullet$ In the UV, the position and FWHM of the 2175 \AA\/ bump are
uncorrelated, consistent with the results of Fitzpatrick \& Massa
(1986).  See Figure 7.  The median value of the bump position for the
core sample 4.59015 $\mu$m$^{-1}$ (2178.6 \AA) with a sample standard
deviation of 0.020 $\mu$m$^{-1}$.  The median value of the FWHM is
0.92 $\mu$m$^{-1}$ with a sample standard deviation of 0.11
$\mu$m$^{-1}$.
\begin{figure}
\plotone{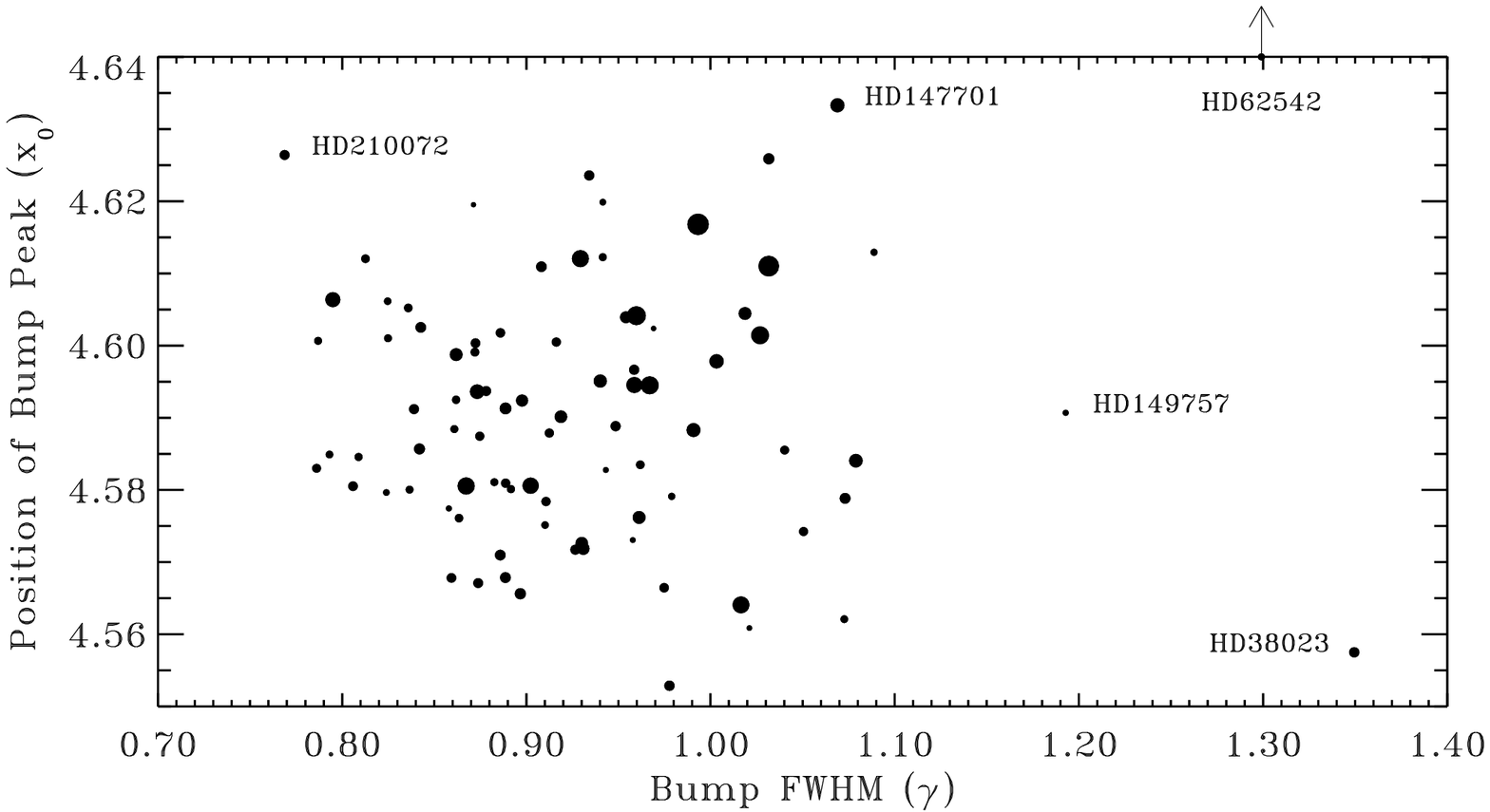}
\caption{Plot of the FWHM $\gamma$ of the 2175 \AA\/ bump versus the
position of the bump peak $x_0$, following the notation of Fitzpatrick
\& Massa 1990.  Both quantities are in units of inverse microns, i.e.,
$\mu$m$^{-1}$. The size of the plot symbols in this and subsequent
figures scales with E(B-V).  Typical internal measurement errors are
$\pm$0.006 $\mu$m$^{-1}$ for $x_0$ and $\pm$0.035 $\mu$m$^{-1}$ for
$\gamma$.}
\end{figure}
\begin{figure}
\plotone{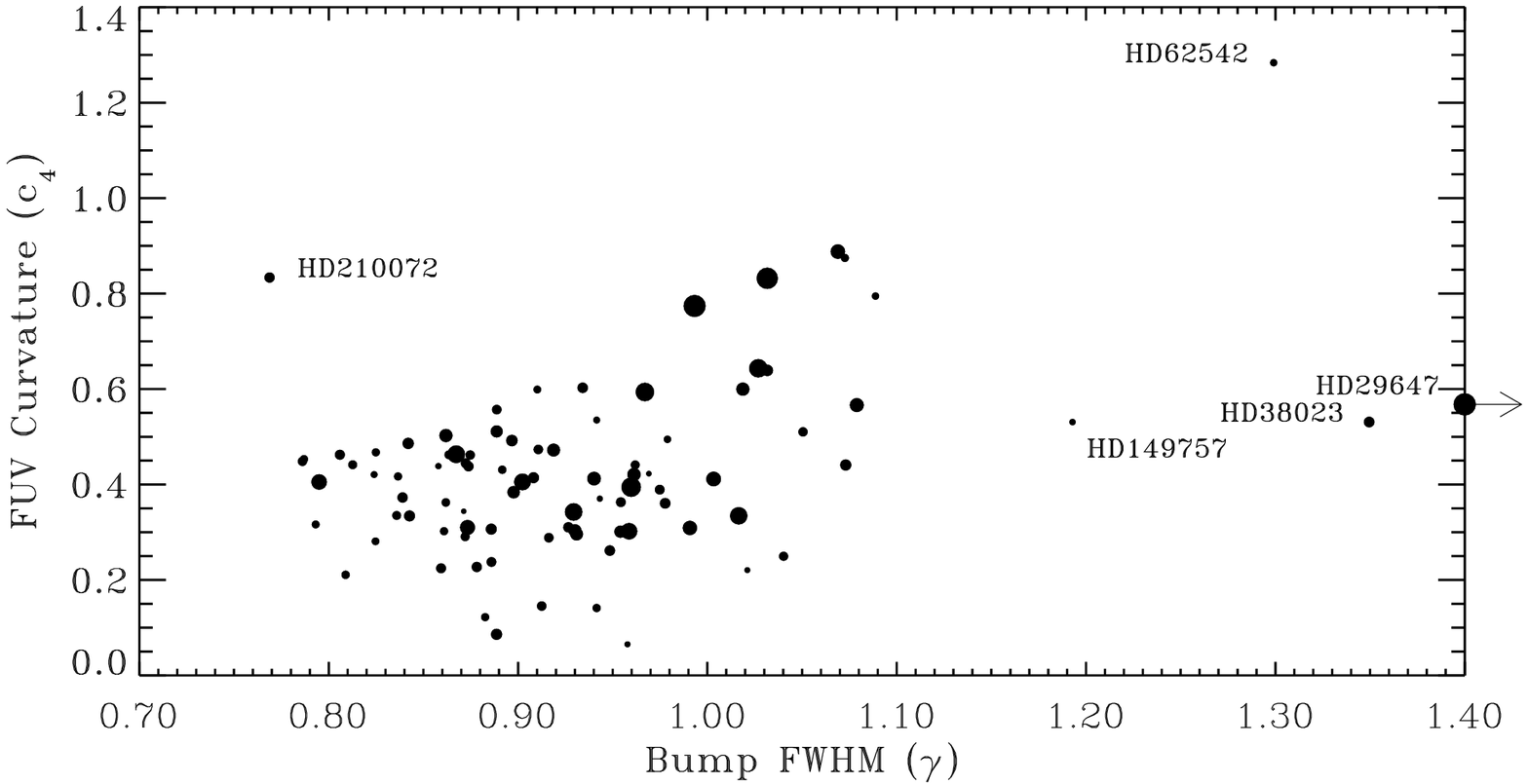}
\caption{Plot of the FWHM $\gamma$ of the 2175 \AA\/ bump versus the
strength $c_4$ of the far-UV curvature.  The typical internal
measurement errors for $c_4$ are $\pm$0.04.}
\end{figure}

$\bullet$ In the UV, the FWHM of the 2175 \AA\/ bump appears related to
the strength of the nonlinear far-UV curvature, in the sense that broad
bumps tend to be associated with large curvature.  See Figure 8.  This
relationship is consistent with that presented by Fitzpatrick \& Massa
(1988).

$\bullet$ In the UV, the slope and intercept of the linear extinction
underlying the bump are well-correlated, consistent with the results of
Fitzpatrick \& Massa (1988). See Figure 9.
\begin{figure}
\plotone{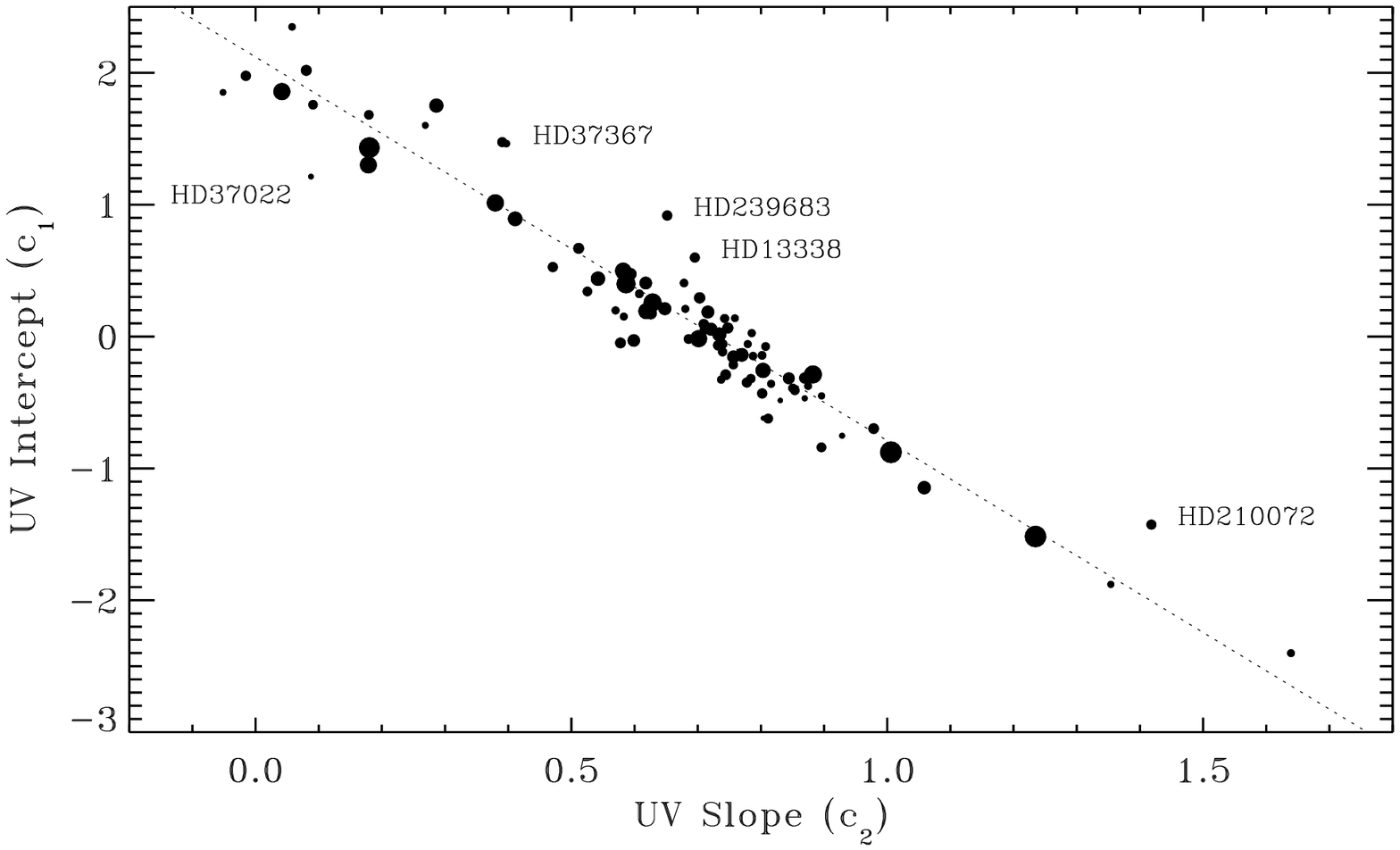} 
\caption{Plot of the slope of the UV linear extinction component $c_2$
versus its intercept $c_1$.  Typical internal measurements errors
are $\pm$0.024 for $c_2$ and $\pm$0.11 for $c_1$. The dotted line is a
linear least-squares fit corresponding to $c_1 = 2.18 - 2.91c_2$.}
\end{figure}

$\bullet$ The slope of the UV linear component and the value of $R_V$
are related: curves with large $R_V$ tend to have flat UV slopes and
curves with small $R_V$ tend to have steep slopes.  See the lefthand
panel in Figure 10. This is the basis of the relationship found by
Cardelli et al. (1989) between UV and optical/IR extinction.  Note that
plots of $R_V$ directly against the level of extinction in the UV,
e.g., ${R_V}^{-1}$ versus {\em E(1250--V)/E(B--V)}, look essentially
identical to the lefthand panel in Figure 10, exhibiting the same
degree of scatter.
\begin{figure}
\plottwo{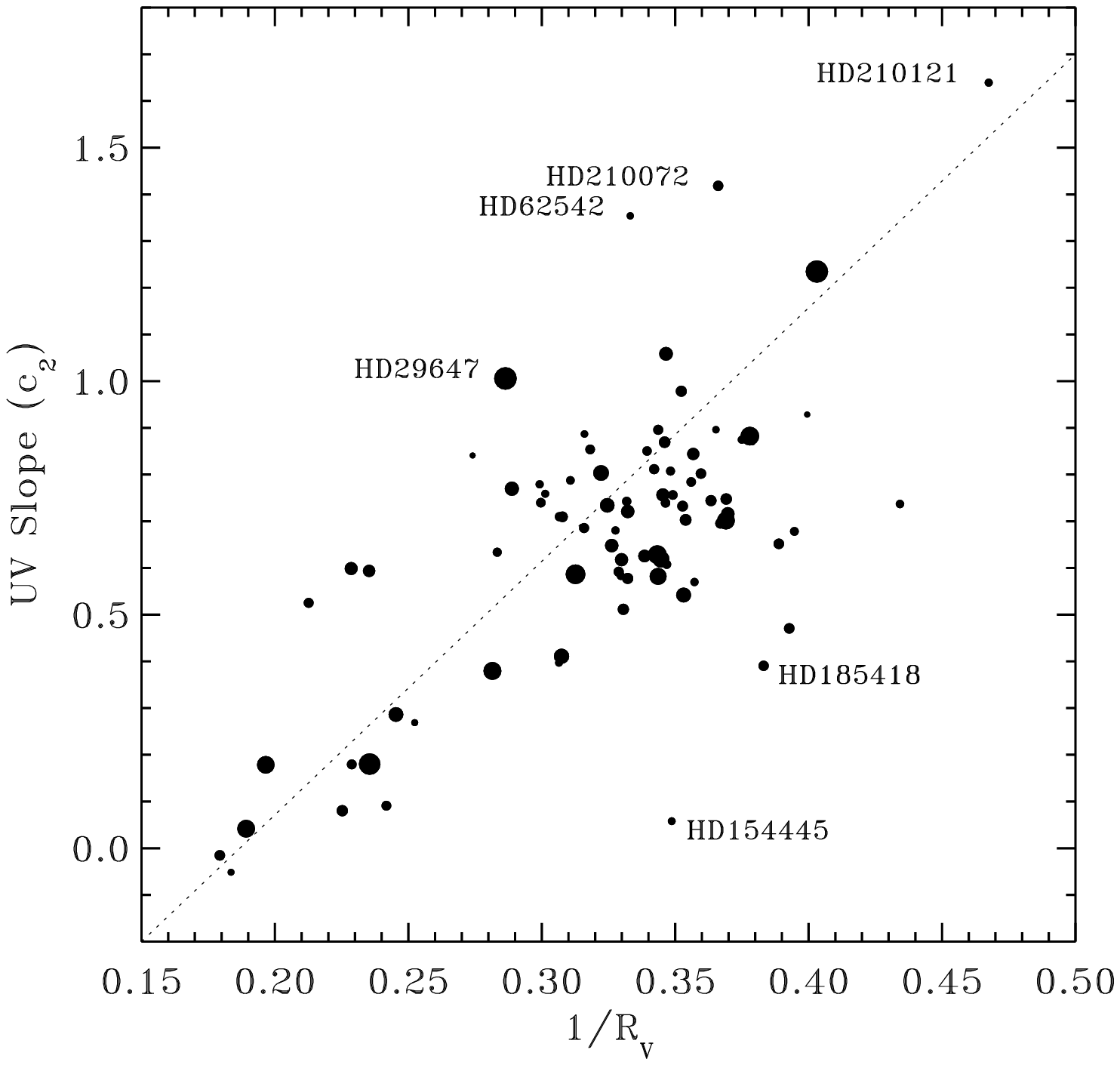}{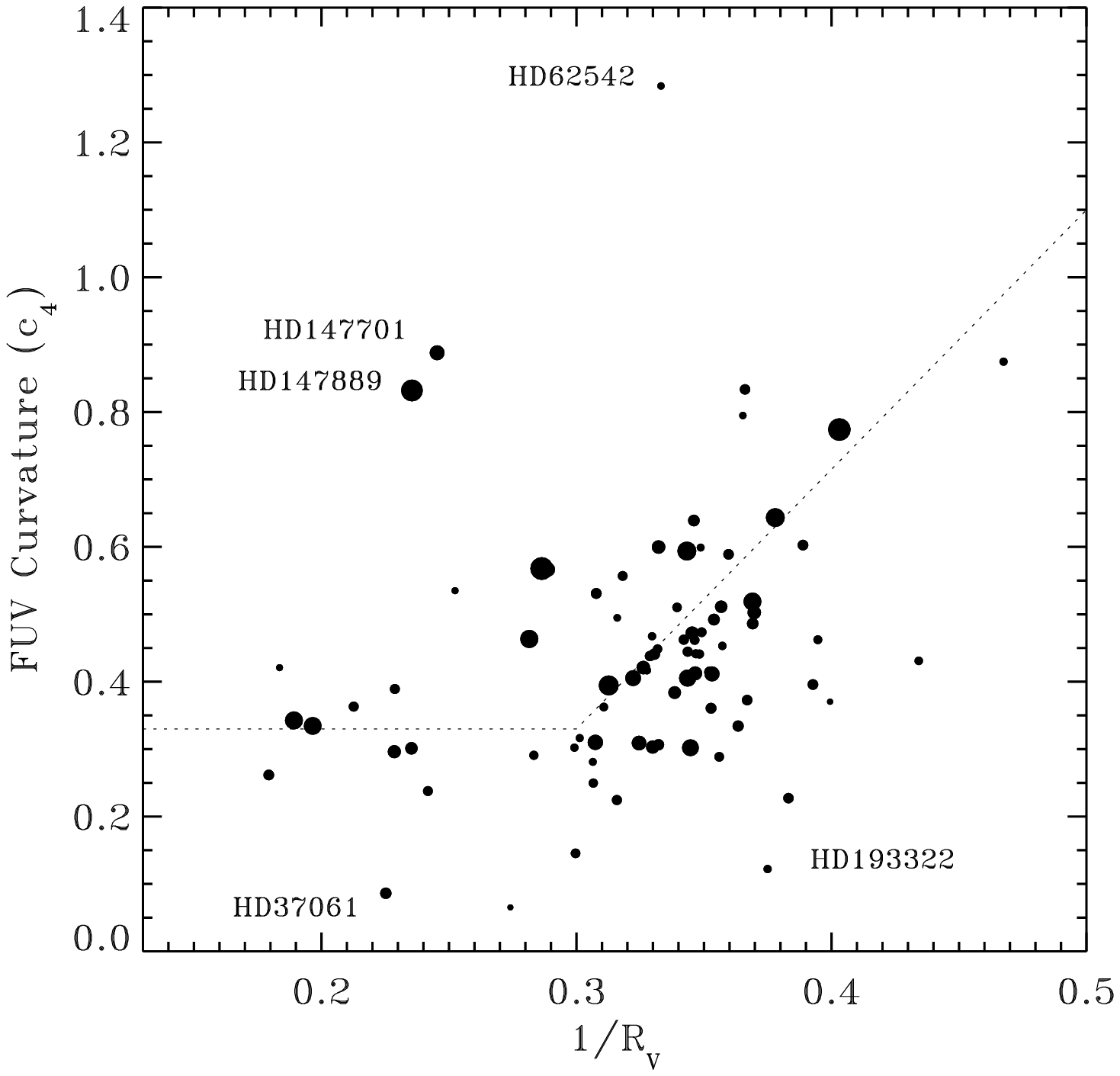}
\caption{Plots of ${R_V}^{-1}$ versus the slope $c_2$ of the UV linear
extinction component (lefthand panel) and the strength $c_4$ of the
far-UV curvature (right panel). The dotted lines in both panels are
{\em not} formal fits to the data and are intended only to draw the eye
to the general relationships between the quantities.}
\end{figure}
\begin{figure}
\plotone{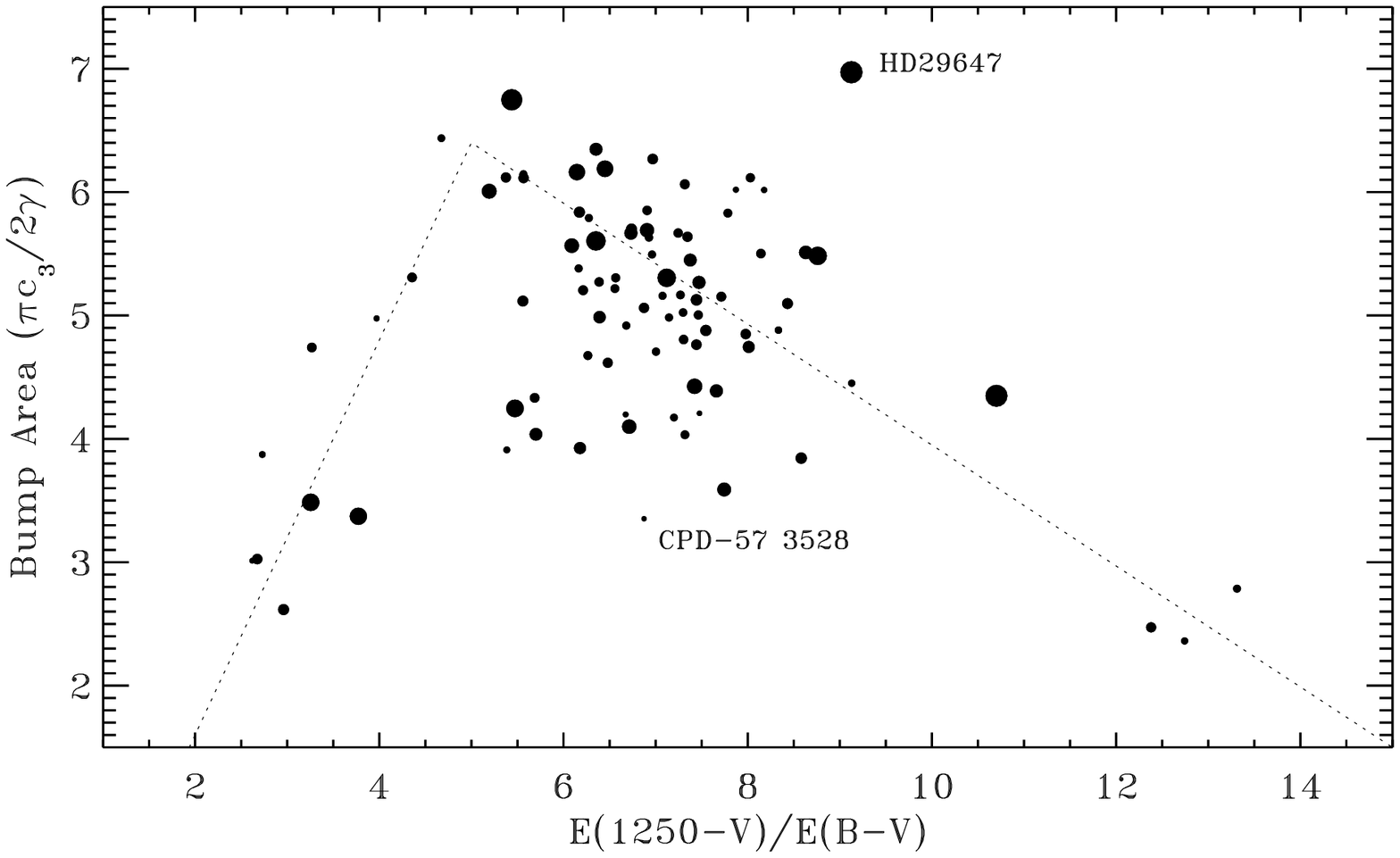}
\caption{Plot of far-UV extinction level {\em E(1250--V)/E(B--V)}
versus the area of the 2175 \AA\/ bump in {\em E($\lambda$--V)/E(B--V)}
extinction curves.  The dotted line is intended only to illustrate
the general relationship between the quantities.}
\end{figure}

$\bullet$ There is a weak relationship between the strength of the
far-UV curvature and the value of $R_V$: for $R_V$ $<$ $\sim$3.3 the
curvature tends to increase with decreasing $R_V$, for $R_V$ $>$
$\sim$3.3 there is no apparent trend. See the right panel in Figure
10.
  
$\bullet$ The strength of the 2175 \AA\/ bump, as measured in curves 
normalized by {\em E(B--V)}, varies over a wide range, but
displays a distinct trend:  the strongest bumps are found for curves
with intermediate levels of far-UV extinction and the bump weakens
progressively as the far-UV extinction strays in either direction
from the intermediate values.  See Figure 11.  Since the level of
far-UV extinction is related to $R_V$, this means that, in general,
weak bumps are found along high-$R_V$ and low-$R_V$ sightlines, while
strong bumps occur along medium-$R_V$ sightlines.

The reader has undoubtedly noticed that the relationships among the
various quantities in Figures 6 through 11 are not particularly
impressive!  The only true correlations are those between $k$ and $R_V$
in the optical/IR domain and between $c_1$ and $c_2$ in the UV domain.
The trends seen among the other quantities are ill-defined ---
although, I believe, undoubtedly real --- and characterized by an
intrinsic scatter much larger than observational errors.  

My primary conclusion derived from these data is that IR-through-UV
Galactic extinction curves {\em should not be considered as a simple
1-parameter family}, whether characterized by $R_V$ or any other
quantity.  While it is straightforward to create a family of curves
which follow the trends established in Figures 6 through 11 --- and
this family is shown in Figures 12 and 13 --- there are actually very
few ``real'' extinction curves which agree in detail with these
idealized forms.  This follows, but extends, the conclusion of Cardelli
et al. (1989) that real deviations from the ``mean law'' exist --- most
curves actually deviate from the mean!  Twenty years ago Massa et al.
(1983) made the comment that ``peculiar extinction is common,''
referring to the wide range of observed curve types.  I think this
statement should be amended to read ``normal extinction is common,''
with the understanding that ``normal'' actually covers a lot of
territory!  The truly peculiar curves, i.e., those which really stand
out from the crowd in Figures 6 through 11, are actually rather rare.

\begin{figure}
\plotone{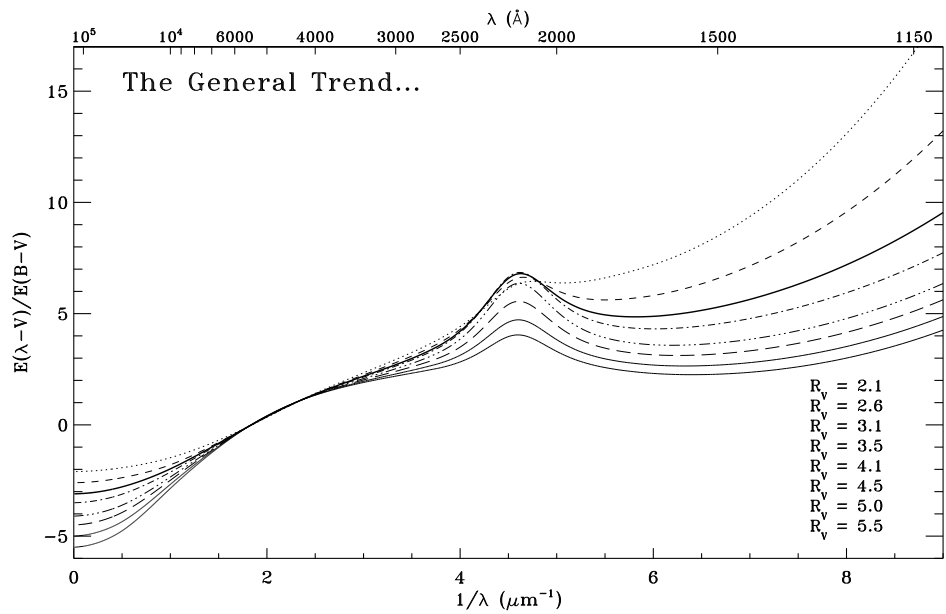}
\caption{A set of normalized extinction curves, ranging from $R_V$ =
2.1 to 5.5, which follow the general trends indicated by the dotted
lines in Figures 6 through 11.}
\end{figure}
\begin{figure}
\plotone{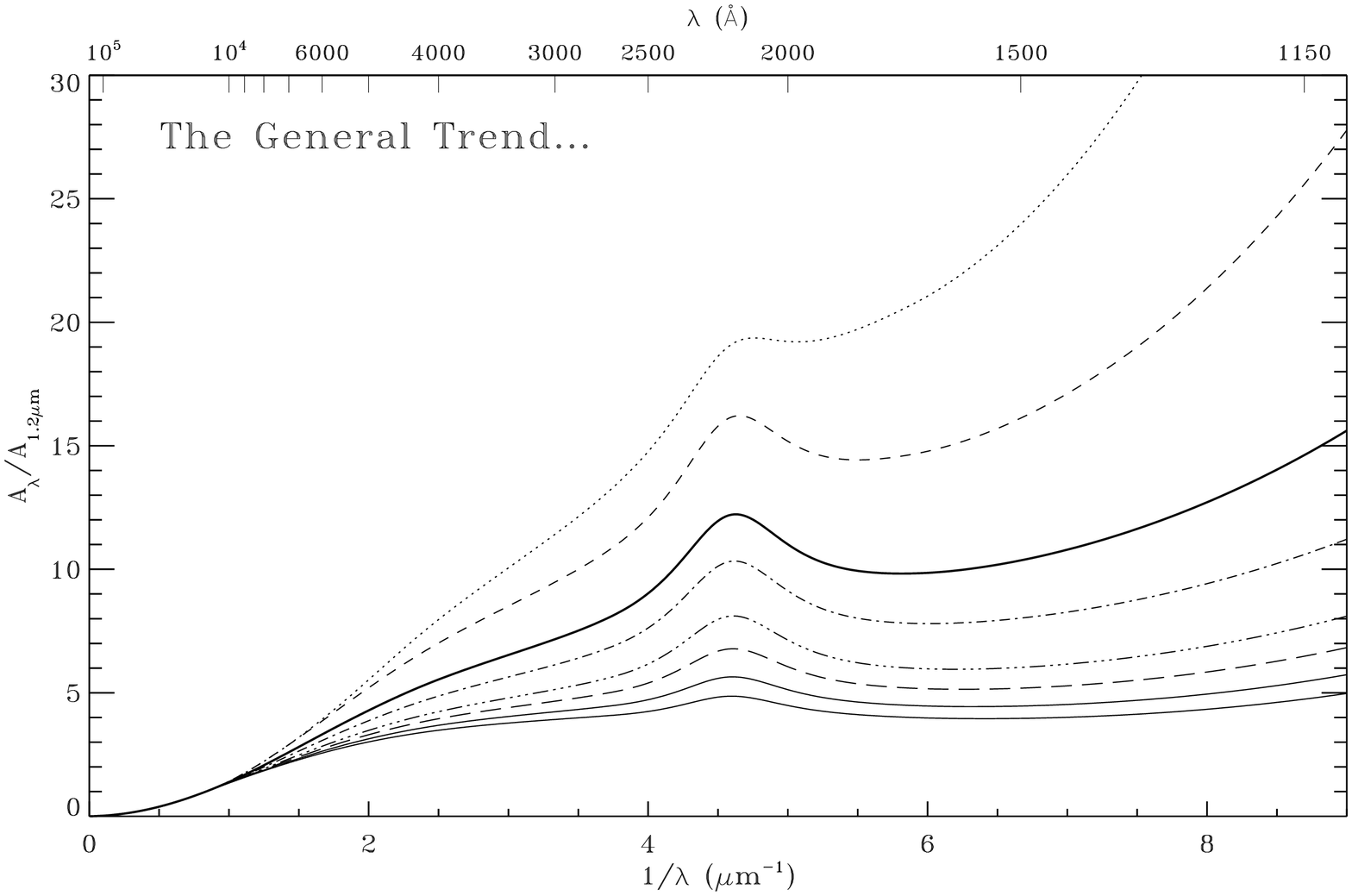}
\caption{The same set of extinction curves as shown in Figure 12, but
presented as total extinction $A_\lambda$ normalized by
$A_{1.2{\mu}m}$, the total extinction at 1.2 $\mu$m.}
\end{figure}

\subsection{The Interpretation}

The data presented above raise a number of interesting issues which, as
an observer, I feel no obligation to either interpret or understand.
However there are a small number of rather obvious points and
suggestions which might help illuminate some of the trends seen.\\

$\bullet$ {\em Size Distribution Matters...} The basic structure and
the multiwavelength coherence seen in the IR-through-UV curves are
determined primarily by 1) the IR power law (Figure 6), 2) the UV
linear slope vs. linear intercept correlation (Figure 9), and 3) the
relation between $R_V$ and the UV slope (Figure 10).  These three
combine such that --- as can be seen clearly in Figure 13 ---
large-$R_V$ curves roll over in the optical and flatten out in the
near-UV while small-$R_V$ curves remain steep throughout the optical
and into the UV. Since basic theoretical expectations (e.g., see
Spitzer 1978, Fig. 7.1) show that extinction becomes grey in the regime
where grains are large compared to the photon wavelengths, the
observations imply that the connection between UV and IR extinction
first noted by Cardelli et al. (1988, 1989) is simply a product of the
grain size distribution. The large-$R_V$ curves reflect sightlines with
larger-than-typical mean grain sizes, while the small-$R_V$ sightlines
reflect smaller-than-typical mean grain sizes.  If the relationship
between $R_V$ and UV extinction were a tight correlation, then we might
be in the uncomfortable position of having to explain a remarkable
uniformity in the form of the grain size distribution among widely
varying sightlines.  However, the observed scatter in Figure 10
probably indicates that Nature actually varies both the high- and
low-end size cutoffs and the shape of the grain size distribution,
producing a range in UV slopes for the same value of $R_V$, and a range
of $R_V$'s for the same UV slope.\\

$\bullet$ {\em The Weak Bumps I...} The trend of weakening 2175 \AA\/
bump with increasing normalized far-UV extinction (and, thus,
decreasing $R_V$) seen in Figure 11 might also be attributable to the
grain size distribution.  Because of the intrinsic steepness of their
extinction in the optical region, grain populations which produce the
high far-UV (low $R_V$) curves are very efficient producers of the
optical color excess {\em E(B--V)}. In addition, such populations have
a larger mean ratio of grain surface area (which is important for
producing extinction) to grain mass, due to their smaller-than-average
size distribution.  As a result of these two effects, the grain
populations which produce low $R_V$ extinction probably have a much
larger ratio of {\em E(B--V)} per unit grain mass than high $R_V$
populations. Since we normalize the curves by {\em E(B--V)} it is
possible that the grain populations which produce the steep,
``weak-bumped'' curves could actually have the same average bump
strength per unit grain mass as in other sightlines. I.e., the apparent
weakness of the bump could be an artifact of the normalization. Testing
this will require measuring the bump strength relative to some absolute
measure of the amount of grain material present along a line of sight,
rather than to {\em E(B--V)}. \\

$\bullet$ {\em The Weak Bumps II...} The weakening of the bump in the
flattest (highest $R_V$) extinction curves requires a different
explanation than suggested above for the steepest curves. The highest
$R_V$ curves in the current sample are all associated with sightlines
close to the Orion Trapezium region or towards the young O star
Herschel 36.  The extinction-producing material is likely freshly
exposed to the interstellar radiation field, having only recently been
deeply embedded in star-forming molecular clouds.  By virtue of their
opacity, we know essentially nothing about the UV extinction properties
in such regions. It is not possible to say if the weak bumps observed
in the large-$R_V$ sightlines are typical of the opaque molecular
regions, or are a result of some specific physical processes operating
in their environments.\\

$\bullet$ {\em Beyond the Milky Way...} Having had a long-standing
interest in the problem, I cannot resist violating the main constraint
on this article and commenting on extinction outside the Milky Way
galaxy --- in the Magellanic Clouds, to be specific.  The steep,
weak-bumped extinction curves of the Large Cloud (LMC) and the really
steep, really weak-bumped curves of the Small Cloud (SMC) have been a
puzzle for 20 years.  Observationally, the situation hasn't changed
much (qualitatively, at least) since I talked about these galaxies at
the Santa Clara dust meeting in 1988.  Theoretically, the idea that
these curves are somehow the result of the progressively lower metal
abundances of the LMC and SMC has been on the table --- but unverified
--- since the first observations.
\begin{figure}
\plotone{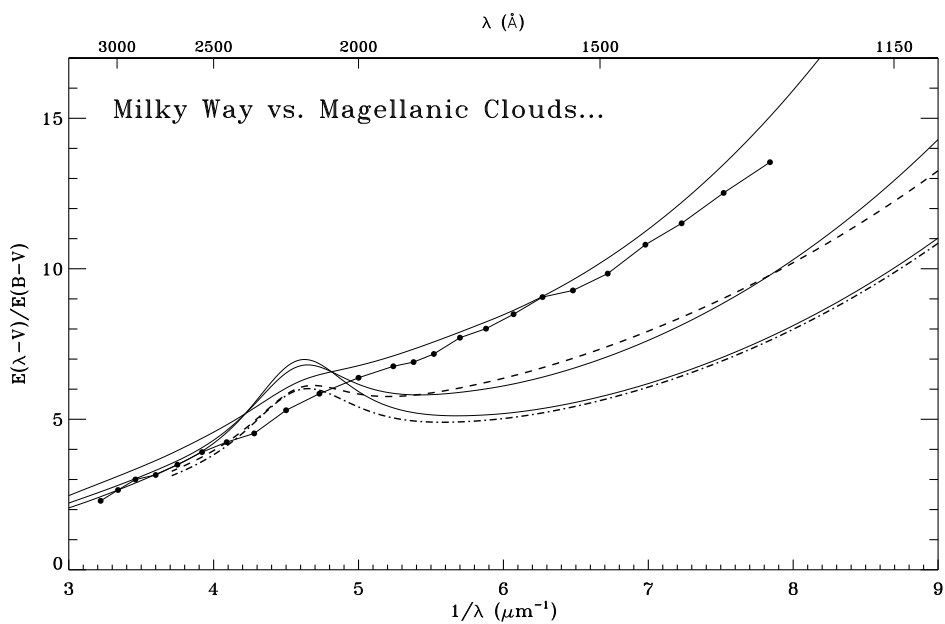}
\caption{A comparison between idealized Galactic extinction curves with $R$ = 1.9, 2.5, and 2.85 and the SMC curve from Pr\'{e}vot et al. (1984; filled circles), the 30 Doradus curve (dashed curve), and the mean LMC curve (dash-dotted curve), respectively.  The latter two curves are from Fitzpatrick (1986), as parametrized by Fitzpatrick \& Massa (1990).}
\end{figure}

Figure 14 shows a comparison between extinction curves from the
Galactic ``family'' featured in Figures 12 and 13 and the Magellanic
Clouds.  The very steep SMC curve has no known Galactic counterparts
but --- as seen in the figure --- it represents only a slight
extrapolation (to $R \simeq 1.9$) of the trends shown in Figures 10 and
11.  The only notable discrepancies in the comparison are in the
strength of the LMC bumps, which are each $\sim$30\% weaker than in the
corresponding Galactic curves.

The results in Figure 14 are very suggestive and lead to an alternate
explanation for Magellanic Cloud extinction: {\em the general
properties of the LMC and SMC curves merely reflect the grain size
distributions along their sightlines and aren't necessarily special
products of alien extra-Galactic environments}.  This is an appealingly
simple idea, and also provides a straightforward explanation for the
presence of sightlines in both galaxies which exhibit Galactic-like
extinction:  these are merely sightlines along which the mean grain
size is similar to that along typical Galactic sightlines.  There may
well be some galaxy-specific details in the curves, perhaps the
slightly weak LMC bumps, but, overall, these curves do not stand out in
a qualitative sense from the Galactic sample.

If this suggestion is correct then the $R_V$ values found in these
galaxies should be low as compared with the Galactic mean.  This is not
obviously the current observational situation (e.g., see the recent
results of Gordon et al. 2003) but I believe that $R_V$ is so difficult
to measure reliably for the lightly-reddened supergiants observed in
the Clouds that the current observations do not at all rule out the
grain-size explanation for Magellanic Cloud extinction. Note that
Rodrigues et al. (1997) have already reached the conclusion, based on
polarization and other data, that SMC grains are smaller than Galactic
grains --- except along the SMC sightline which exhibits Galactic-like
UV extinction.  In addition, a number of authors have modeled
Magellanic Cloud extinction and shown that modified grain size
distributions provide a plausible explanation for the observed curves
(e.g., Pei 1992; Weingartner \& Draine 2001; Clayton et al. 2003a).  The
contribution here is simply to point out that Magellanic Cloud
extinction is consistent in general with trends observed along Galactic
sightlines where the mean grain size is probably
smaller-than-average.\\

$\bullet$ {\em Maybe It's Better Than It Looks...} Finally, it is
possible that a number of effects conspire to make the various
extinction properties appear less correlated with each other than they
really are.  I have screened the sample for obvious stellar
peculiarities, which might distort the shape of the derived curves, but
more difficult-to-diagnose problems may still exist for some curves.
In addition, significant circumstellar material might be present in the
vicinity of some of the stars, resulting in light scattered into the
line of sight. Both these issues can be addressed by more detailed
examination of the stellar data.  Lastly, there is little doubt that
some of the observed scatter arises because some sightlines traverse
multiple regions of greatly different dust grain properties, e.g., a
$R_V = 2.5$ region and a $R_V = 4.5$ region.  Such superpositions will
distort correlations between extinction properties which are related in
an intrinsically non-linear way, as in Figures 10 and 11, but will not
distort intrinsically linear correlations, as in Figures 6 and 9. For
example, even if the strength of the 2175 \AA\/ bump were perfectly
correlated with the far-UV extinction in the manner suggested by the
dotted diagonal lines in Figure 11, the region between the diagonals
would still be filled-in as a result of composite sightlines.  With our
larger sample of curves, Massa and I will attempt to identify composite
sightlines and see if the some of the extinction properties are
intrinsically better-related that first glance suggests.

\section{Other Extinction Issues}

By using virtually all my allotted space (and available energy!)
concentrating on the issue of the ``1-parameter family'' I have
slighted much recent work and many other researchers working hard on
Galactic extinction-related issues. I cannot correct this fault in the
small amount of remaining space, but do want to conclude by noting some
currently active areas of investigation:

$\bullet$  {\em Far-UV Extinction...} Measurements of extinction in the
1150--912 \AA\/ region are very difficult due the presence of the
Lyman-series absorption lines of H I and the Lyman- and Werner-series
absorption bands of $H_2$.  In addition, heavy line-blanketing in the
spectra of the early-type stars required for extinction measurements
make the results very sensitive to spectral mismatch error.  Recent
results have been published by Hutchings \& Giasson (2001), based on
{\em FUSE} data, and Sasseen et al. (2002) based on {\em ORPHEUS}
data.  At this point, the far-UV extinction appears to be a smooth
extrapolation of the rise seen in the mid-UV range, with no sign of a
turnover.  Much work remains to be done in this wavelength range.

$\bullet$  {\em Extinction Features...} The 2175 \AA\/ bump is the
strongest observed extinction feature.  Other dust-related
spectroscopic features include the optical Diffuse Interstellar Bands
and the IR solid-state emission and absorption bands.  Clayton et al.
(2003b) recently performed a detailed search in the UV for other
extinction-related features, particularly as might be associated with
absorption by polycyclic aromatic hydrocarbons.  They found no evidence
for such features down to a 3$\sigma$ detection limit of 0.02$A_V$.  If
these results (based on two lines of sight) are the norm, then UV
extinction curves are evidently very, very smooth.

The optical and near-IR spectral regions would benefit from a similar
study.  For years there have been tantalizing hints of structure in the
$V$ band region, known as the ``Very Broad Structure'' or ``VBS'' (van
Breda \& Whittet 1982; Bastiaansen 1992) but no definitive
characterization of its properties.  This lack of data on an apparently
observationally-accessible portion of the spectrum is undoubtedly due
the inherent difficulties in obtaining precise ground-based
spectrophotometry.

$\bullet$  {\em Morphology...}  Despite many years of study, our
understanding of extinction and its relationship with dust grain
properties and environments still relies heavily on simple
morphological investigations, e.g., Figures 6--11 of this paper.  Such
studies also include expanding the range of interstellar environments
in which extinction is measured, as well as searching for relationships
between extinction and other measurable quantities in the ISM.  Recent
work along these lines includes the Clayton, Gordon, \& Wolff (2000)
study of low density sightlines in the Milky Way and comparisons
between the extinction properties along specific sightlines and their
molecular content (Burgh et al 2000; Rachford et al. 2002).\\
 
By the time the next interstellar dust meeting rolls around, it is
likely that studies such as these few examples listed above will have
greatly improved our ability to characterize extinction as a function
of interstellar environment and provided valuable insight into the
natures of interstellar grains and the processes which modify them.
 

%
%

\end{document}